# Vehicular Ad-hoc Networks

**Asad Maqsood , Rehanullah Khan**
Department of Electrical Engineering
Sarhad University of Science and IT
Peshawar, KPK, Pakistan
asad.maqsood@gmail.com**,** rehanmarwat1@gmail.com

**Abstract**
*Modern day's vehicles require advanced communication system on board to enable passengers benefit the most from available services. IEEE 802.11p is the new extension of IEEE 802.11 standards; especially proposed for the high vehicular environment. The WAVE documentation represents enhancements to the Media Access Control (MAC) and Physical (PHY) layer of IEEE 802.11 standards to work efficiently in high vehicular environment. In this research work, the main emphasis is on the new IEEE 802.11p enhancement of MAC and PHY layers. More specifically, the target of this research is to setup a simulation environment which will allow us to investigate the use of real time voice application, using IEEE 802.11p (WAVE) enhance setting, in a single hop and multi-hop environment where nodes are not directly connected. Also, the evaluation of transmission between moving nodes are tested by simply sending and receiving FTP file between them with varying speed of the moving nodes.*

## 1. Introduction

In near future, modern vehicles will be equipped with on board intelligent units, which will inform the drivers about a range of safety information, to help them raise their vehicle safety. Apart from safety applications, the commuters will also be able to enjoy the non-safety application e.g. on board internet surfing, multimedia applications and so on. Currently, several ideas around the world are considering vehicular safety applications by means of short range wireless communications. One of the major improvements in vehicular communication is from the Federal Communication Commission (FCC), allocated 75 MHz spectrum at 5.9 GHz for intelligent transport system (ITS) application in 1999 in United States. By adopting the Dedicated Short Range Communication (DSRC) technology, multi-hop ad hoc will become the mainstream technology in modern vehicular environments. In Vehicular Ad-hoc Network (VANET), the vehicles should be able to communicate locally without the need of any centrally managed infrastructure or base stations controlling the medium access. For multi-hop communication, in VANET, data is forwarded to the destination vehicle by using location-based ad hoc routing protocol instead of IP addresses. There are range of multi-hop routing protocols developed over the years e.g. AODV, DSR, OLR are the few of them. The VANET plays an important role in the development of Vehicular-centered applications where cars collect the local information about the road conditions distribute this information locally and overwhelm local information from the nearby vehicles. Apart from the safety information the non-safety information are also provided to the commuters, for this purpose the Internet Gateways (IGWs) are installed along the roadside to provide a temporary internet access. Though, mobility management is require to handle the mobility of a vehicle in IGW to ensure that the requested data is from the internet always deliver to the appropriate vehicle through IGW. The vehicle must also be able to discover the IGW within VANET even its multi-hop away. To address these problems the IEEE 802.11p task group made some enhancement to the MAC layer for better support of safety and non-safety applications and PHY layer to support communication distance up to 1000m. Also to enhance larger distance, multi-hop communication is to be supported in efficient way. By enforcing such a technology in transportation, congestion problems could be solved out which could save billions of dollars of fuel, also millions of hours of waste of time on the road.

## 2. IEEE 802.11 Standards

The IEEE 802.11 standard describes the PHY and MAC layers specifications of Wireless LANs. There are several methods for data transmission between two nodes at the physical layer. Nodes can use Direct Sequence Spread Spectrum (DSSS), orthogonal frequency distribution modulation (OFDM), or frequency hopping spread spectrum (FHSS). For access mechanism two different methods are used, Distributed Coordination Function (DCF) and Point coordination Function (PCF). The DCF mechanism uses CSMA/CA for access method. In CSMA/CA the exchange of request-to-send (RTS), clear-to-send (CTS), a data and acknowledgement (ACK) are required for each sending data packet. To avoid collision a back-off mechanism is deployed before the start of transmission. If the channel is free, an additional random time for listening to the channel before starting the transmission is used. This interval is called DIFS (DCF Inter-frame Space). The sender node can start transmission if the channel remains free for DIFS. Contention window (CW) is maintained at each node to determine the amount of time a station should wait before transmitting. The value of CW remains the same each time and ACK is received from the receiver, but it will increase if the transmission fails. The increase in CW will results in an increase of value of random back off timer. Finally, to deal the problem of hidden node, RTS and CTS are use. An RTS frame is sent by the sender before the transmission and CTS frame is sent back by the receiver to inform his availability for receiving the data. In PCF polling methodology is used which allows the point coordinator node to poll different nodes that need to send the data and to which, if they are polled, they send their packets. Also PCF use a contention-free period (CFP) and a contention period (CP). During the CFP, a PCF mechanism is used while in CP a DCF method is used.

The two mechanism used by the WLAN, do not have any room for supporting the real-time traffic since low end-to-end delay or jitter cannot be guaranteed polling or in a CSMA/CA point of view, especially in a VANET environment. IEEE is working on IEEE 802.11p extension, which is an enhanced version of IEEE 802.11standards, designed for VANETs and to support multimedia transmission efficiently in Vehicular environment.

## 3. IEEE 802.11p Enhancements

In this section, we discuss the enhancements related to the IEEE 802.11p standard.

### 3.1 IEEE 802.11p (WAVE) MAC Enhancements

Most of the changes in IEEE 802.11p standard are related to the MAC layer. MAC layer changes are often software base and can be update quite easily rather than PHY layer. The Enhancements in MAC layer in IEEE 802.11p are listed below.

### 3.1.1 WAVE Mode

IEEE 802.11 MAC operation are too time intense, where in high vehicular environment, vehicular safety communications use cases demand instantaneous data exchange capabilities and cannot afford typical 802.11 method of scanning channels for beacon of BSS and execute multiple handshakes for establishment of communication. It is essential for all IEEE 802.11p complaint devices to have radio configured in a same channel with the same BSSID for safety communication with no delay. For example if a vehicle crossing other vehicle in the opposite direction, the time for communication may be extremely short due to the vehicles dynamics. WAVE mode is introduced in IEEE 802.11p WAVE for capability enhancement. In WAVE mode a wildcard value is assign to BSSID for transmitting and receiving of data frames without the need for the node connect to BSS. This is very beneficial for the vehicle communicating for a short interval of time and for safety communication which do not require additional overhead for simple communication, as long as they use the same channel with wildcard BSSID.

### 3.1.2 WAVE BSS

The overhead of typical BSS (Basic Service Set) setup is too much expensive for both safety and non safety applications. A vehicle approaching a road side station that offers, suppose services like local information, it can hardly afford few seconds that are required in typical WLAN connection setup, because due to the dynamics of vehicle the total time it stay in the range will be too short then waiting for connection .Analyzing this factor WAVE standard introduced WBSS (WAVE BSS), which is the enhancement of BSS type. In WBSS environment, an STA forms a WBSS by first transmitting an on demand beacon. The WAVE station uses that demand beacon, which uses the well known beacon frame and needs not to be repeated every so often, to advertise a WAVE BSS unlike BSS. Upper layer mechanism above the IEEE 802.11 creates and consumes such advertisements. It contains all the necessary information need by the receiver station to understand the services offered in the WBSS in order to decide whether to join the WBSS and if needed configure itself into a member of the WBSS. In other words if station decides to join will need only WAVE advertisement for complete joining process with no further overhead.

### 3.1.3 Wildcard BSSID Usage

The 802.11p WAVE was suggested for safety as a key, the use of wildcard BSSID is supported for stations even they are already belongs to WBSS (i.e. configured with a particular BSSID). Means an STA in WBSS will be still in WAVE mode in order to transmit frames with wildcard BSSID in order to reach its neighbour STAs in cases of safety concerns. Also, an STA already in a WBSS and having its BSSID configured for filtering accordingly can still receive frames from STA's outside the WBSS with wildcard BSSID. The main purpose of BSSID configured with wildcards strengthen the sending and receiving data frames for safety communication but also support signaling of future upper layer protocols in Ad-hoc environment.

**3.1.4 Distributed Services**

The Wave complaint device still support Distributed services. In WAVE BSS the concept of wildcard is used to send and receive data frames, which introduces complications. It is more probable that a radio will be restricted to send a data frame with the wildcard BSSID on if the "To DS" and "From DS" bits are set to 0. Means radios communicating in WAVE BSS environment should send data frames to known BSSID for accessing the DS.

**3.2 IEEE 802.11p (Wave) PHY Enhancements**

In IEEE 802.11p standards, the PHY layer is not change as such because the 802.11a radios are already operate at 5 GHz and it is not difficult to change the configuration of these ratios to work on 5.9 GHz band in U.S and internationally. The purpose of minimum changes to IEEE 802.11 PHY is that WAVE device can communicate efficiently among different fast moving vehicles in the highway environment. On the other hand MAC layer enhancements are basically software updates bases, which is also easy to make, while PHY level enhancement minimal in order to shun designing an entirely new architecture for wireless technology. The few enhancement made are given below.

**3.2.1 Channels Frequency**

IEEE 802.11p is based on IEEE 802.11a which basically use OFDM modulation scheme for communication with 20 MHz channels, while in IEEE 802.11p 10 MHz channel is used. The implementation of 10 MHz channels is straightforward since it mainly about doubles all of OFDM timing parameters used in the regular 20 MHz 802.11a transmissions. The main reason for this scaling of 802.11a is to describe the increased RMS delay spread in the vehicular environment. For accommodating the large communication rage in vehicular environment, four classes of maximum allowable Effective Isotropic Radiation Power (EIRP) up to 30 W (44.8dbm) are allocated in IEEE 802.11p. For emergency vehicles approaching the highest values is reserved which is typically 33dBm for safety relevant messages.

**3.2.2 Enhanced Receiver Performance**

One of the problems which is well known and is natural property of wireless communications is cross channel interference. In U.S and (expectedly) internationally there are number of channels available for 802.11p deployment and usage. There is increase concern about cross channel interference among closely distributed vehicle on the road. The measurement presented by (Rai, v., Bai, F., et al.2007).demonstrates the potential for immediate neighbouring vehicles to interfere with each other's communication. If they are using two adjacent channel, for example a vehicle using channel 176 , could be interfered by the vehicle adjacent to it using channel 178, for receiving safety message send by another car ahead. However, IEEE 802.11p introduces some improvement in receiver's performance, required in adjacent channel rejections. There are two categories of requirement listed in the proposed standards. The first category is mandatory and to be understood generally to be reachable with today's chip manufacturers. The second category is more stringent and optional.

## 4. Simulation Setup

The scenarios were built using the IEEE 802.11p standards in the simulators. On the physical layer the most robust PHY mode is chosen (Binary Phase Shift Keying with 50% redundancy, BPSK1/2). The transmission power is 30dBm (1W). Omni-directional antennas are used, so no antenna gain is involved. The center frequency is 5.9GHz with a channel bandwidth of 10MHz, with the 6Mpbs data transmit rate.

**4.1 Scenario Description**

The scenario chosen for the evaluation is the highway scenario. In the first scenario a single hop communication between the mobile node and the roadside server is represented. The first scenario consist three different speed limits for the mobile nodes. First the node move with the 32 km/h (20mi/h), then the same node with the same environment and setting only the speed is alter to 65 km/hr (40mi/h), and 97 km (60mi/h). The second scenario is the same as the first one but a second server is planted to check the mobile node communication with that server multi-hop away. In the third scenario we have two mobile nodes in the same direction. The first node is moving with 32km/h (20mi/h), while the second node's speed changes. At first the second node move with 32km/h(20mi/h), then 65 km/h(40mi/h) and 97km/h(60mi/h) respectively. The communications between the nodes are tested using the FTP services. The IEEE 802.11p protocol stack, the channel and mobility model were implemented in our discrete event simulator OPNET$^{TM}$ modeler.

## 5. Simulation Results

In this section, we present the simulation results in different configuration scenarios.

## 5.1 Single Hop Communication

The simulation results are presented next in Figures 1-2. The simulation for single hop communication consists of three simulations. The moments of car was 32 km/h (20mi/h), 65km/h (40mi/h) and 97km/h (60 mi/h) respectively. The graph represents the comparative average wireless through and average delay put in figures 1-2 of nodes in three different speed ranges.

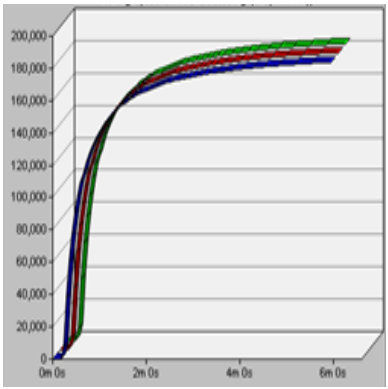

Figure 1: Average through put

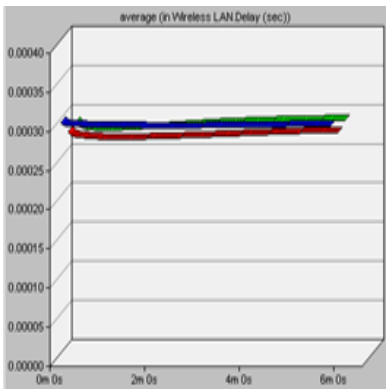

Figure 2: WLAN delay (sec)

AODV routing protocol is used for the routing purpose. The average delay recorded of the route discovery is presented in figure 3. As one can see it is the same in different speed scenario. Also for signaling purposes the H323 protocol is used for registration, call admission control, and call signaling. In figure 4 the call admission and registration time is shown in three dimensions graph to so the impact of the vehicle density and speed on each of the simulated matrices can be easily evaluated.

In next graphs shown in figure 5-6, the performance of AODV routing protocols is shown. The average packets send and received in all different speed scenarios from sender and receiver are about the same.

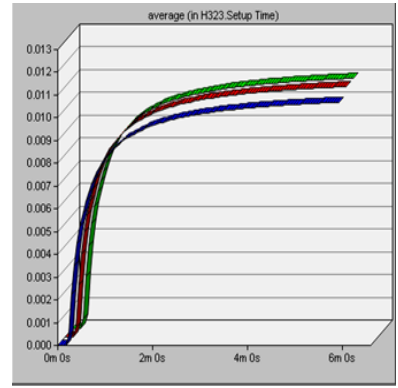

Figure 3: AODV route discovery time

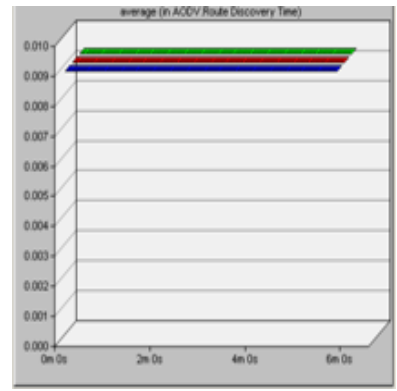

Figure 4: H323 call setup time

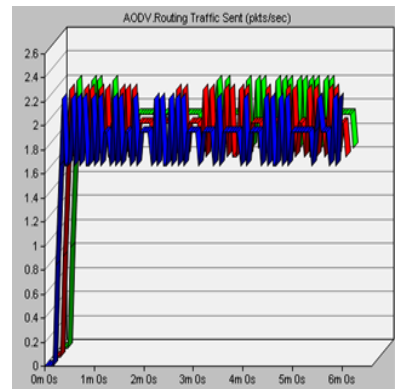

Figure 5: AODV routing traffic sent

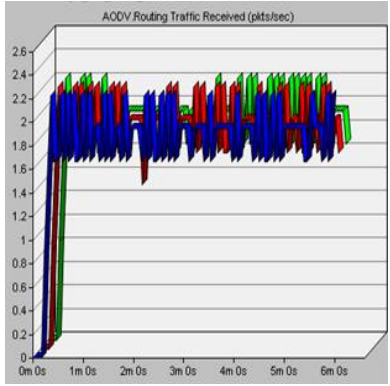

Figure 6: AODV routing traffic received

### 5.2 Second Scenario Multiple Hops

In the second scenario, the whole system is kept the same except a new fixed server is introduced to extend the limit of moving node. The second server is kept in such distance that it is not interrupted with the channels of the first server. The second server has the same properties as first. In this evaluation, we tried to communicate the moving node with the server a hope away. The figures 7-8 shows the throughput and voice packets send/receive among nodes and servers. Figure 9-10 represents the AODV route discovery and H323 call setup time.

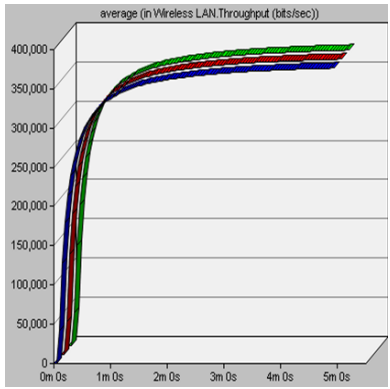

Figure 7: Wireless LAN throughput

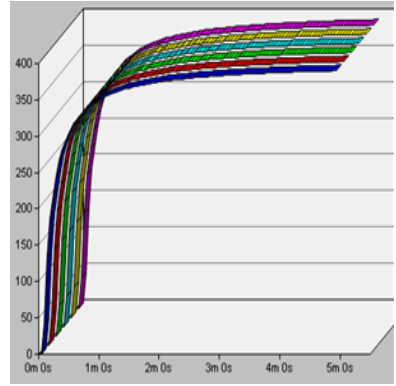

Figure 8: Packets transmission

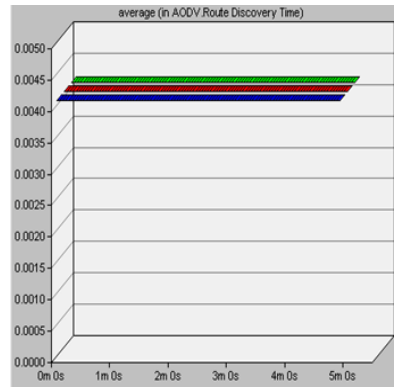

Figure 9: AODV route discovery time

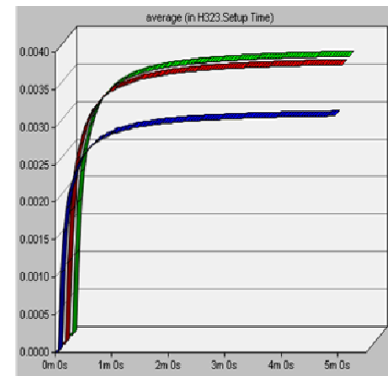

Figure 10: H323 call set up time

Figure 11 represent the average end-to-end delay of the voice packet send across the multi-hop. The multi-hop communication is carried out with the server not in the direct range of the moving range. This is about 0.06 seconds. While WLAN delay is about 0.0003seconds represented by figure 12.

Figure 13-14 represent the AODV routing traffic (packet/sec) sent and received. The sent packets are about three packets per second and receive traffic about six packets per second. It shows the amounts of received packets are from both the servers. In this graph the mobile node efficiently communicating with the server not in the direct range

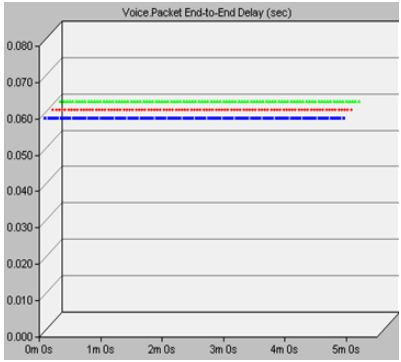

Figure 11: End-to-end voice packet delay

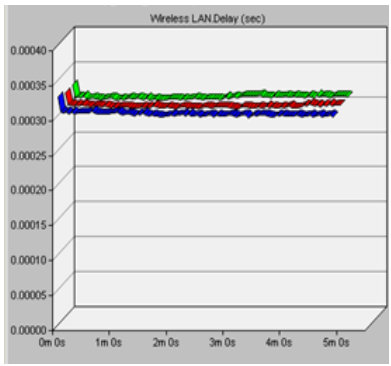

Figure 12: Average WLAN delay

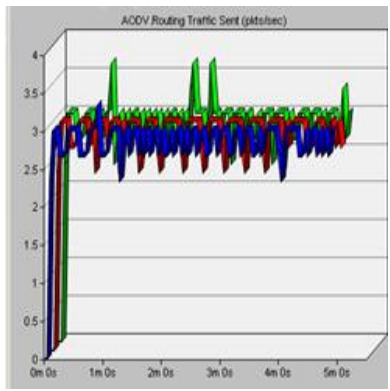

Figure 13: AODV routing packet sent

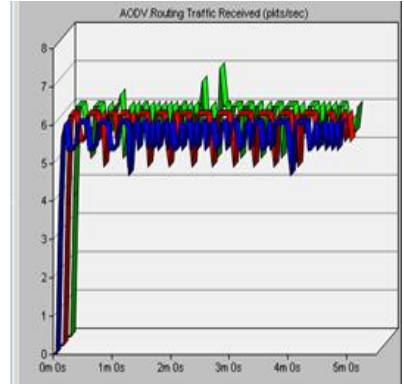

Figure 14: AODV routing packets received

### 5.3 Mobile Node to Node Communication

The third scenario is based on node to node communication while on both the nodes are on the move. This kind of communication is typical Ad-hoc communication, as it is direct communication between the communicating nodes. A FTP file transfer is chosen for file upload or download operation. Figure 15-16 represent the Wireless LAN through put and average delay occur between the moving nodes. The scenario where both cars moving with 32 km/h the average wireless delay is low because of the equal speed, then the nodes moving with 65 km/h and 97 km/h respectively. In fig 16 the delay variation fluctuates so often because of the dynamics of the moving vehicle, but as a whole the delay is still very low and affordable in vehicular environment.

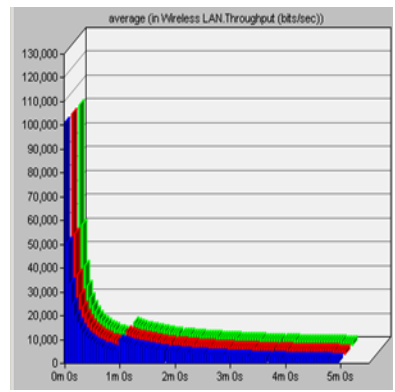

Figure 15: Average WLAN throughput

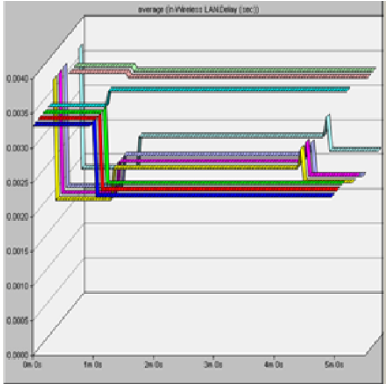

Figure 16: Average WLAN delay

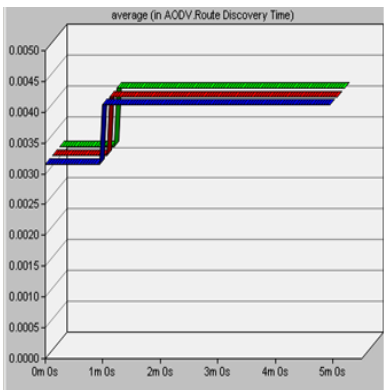

Figure 17: AODV route discovery

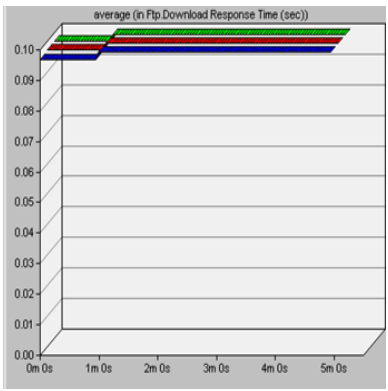

Figure 18: FTP downloads response

Figure 17 represent the average AODV routing protocol route discovery. Figure 18 represents the Average response from the FTP server to download or upload an FTP file.

## 6. Conclusion

In this paper, the enhancement of IEEE 802.11p (WAVE) is discussed. To investigate its impact on the single hop and multi-hop communication in high vehicular environment, real time voice applications are used. AODV is used as a multi-hop routing protocol and H232 codec for voice applications. Different parameters were checked during the simulation regarding the quality of communication. To investigate the communication between the neighboring nodes, FTP services are used. The results show that WAVE complaint applications and devices can greatly improve the communication range and performance of VANET, by supporting efficient multi-hop communication and reducing delay and connection time.